\documentclass[12pt]{article}
\usepackage{amsmath}
\usepackage{amssymb}
\usepackage{amsfonts}

\renewcommand{\theequation}{\arabic{section}.\arabic{equation}}

\title{
Deformed oscillator algebra approach of some quantum superintegrable Lissajous systems on the sphere and of  their rational extensions}

\author{Ian Marquette$^{1,}$\thanks{Electronic address: i.marquette@uq.edu.au} \ and Christiane Quesne$^{2,}$\thanks{Electronic address: cquesne@ulb.ac.be}\\
{\small\sl $^1$ School of Mathematics and Physics, The University of Queensland,}\\
{\small \sl Brisbane, QLD 4072, Australia}\\
{\small\sl $^2$ Physique Nucl\'eaire Th\'eorique et Physique Math\'ematique, 
Universit\'e Libre de Bruxelles,} \\ 
{\small \sl Campus de la Plaine CP229, Boulevard~du Triomphe, B-1050
Brussels, Belgium}}
\date{ }
\begin{document}
\baselineskip=22pt plus 1pt minus 1pt
\maketitle

\begin{abstract}
We extend the construction of 2D superintegrable Hamiltonians with separation of variables in spherical coordinates using combinations of shift, ladder, and supercharge operators to models involving rational extensions of the two-parameter Lissajous systems on the sphere. These new families of superintegrable systems with integrals of arbitrary order are connected with Jacobi exceptional orthogonal polynomials (EOP) of type I (or II) and supersymmetric quantum mechanics (SUSYQM). Moreover, we present an algebraic derivation of the degenerate energy spectrum for the one- and two-parameter Lissajous systems and the rationally extended models. These results are based on finitely generated polynomial algebras, Casimir operators, realizations as deformed oscillator algebras and finite-dimensional unitary representations. Such results have only been established so far for 2D superintegrable systems separable in Cartesian coordinates, which are related to a class of polynomial algebras that display a simpler structure. We also point out how the structure function of these deformed oscillator algebras is directly related with the generalized Heisenberg algebras (GHA) spanned by the nonpolynomial integrals.
\end{abstract}

\vspace{0.5cm}

\noindent
{\sl Runnning title: Deformed oscillator algebra}


 
\newpage
%
%
\section{INTRODUCTION}

Realizations of polynomial associative algebras, arising in the classification of two-dimensional quantum superintegrable systems, in terms of a deformed oscillator algebra have proved very useful to study representations and algebraically obtain the energy spectrum of such systems. Initiated by Daskaloyannis \cite{daska} in the case of quadratic associative algebras generated by two second-order integrals of motion, they were later on extended to cubic \cite{marquette09} and quartic \cite{marquette13a} associative algebras. Very recently, the polynomial associative algebras generated by a second-order integral of motion and one of order $M$ were studied in their full generality \cite{isaac}.\par
%
%
These realizations have among others found applications to new superintegrable systems connected with exceptional orthogonal polynomial (EOP) families \cite{marquette13b, marquette13c}, a domain that has aroused a lot of interest (see, e.g., \cite{cq08, cq09, odake09, grandati11, gomez12, odake11, grandati12, odake13, grandati13, gomez14} and references quoted therein) since the introduction of such families \cite{gomez09, gomez10}. In this context, some other procedures, such as the recurrence relation method \cite{kalnins} and the direct action of the ladder operators, used to build the integrals of motion \cite{marquette10}, on the constituent one-dimensional Hamiltonian eigenstates have also been used \cite{post, marquette14}.\par
%
%
All our previous studies on this subject \cite{marquette13b, marquette13c, marquette14} were devoted to superintegrable systems separable in Cartesian coordinates. We will start considering here other types of systems and turn our attention to a recent study of so-called (one- and two-parameter) Lissajous systems on the sphere \cite{calzada14a, calzada14b}. The name ``Lissajous'' given to such systems comes from the fact that the corresponding classical trajectories on the sphere look like the well-known Lissajous curves on the plane. The authors of Refs.~\cite{calzada14a, calzada14b} used a unified procedure to compute the integrals of motion of the classical and quantum superintegrable systems, based on the factorization method. Furthermore, and more importantly for us, they devised a procedure to construct a set of polynomial integrals of motion equivalent to the set of nonpolynomial ones so derived.\par
%
%
The purpose of the present paper is twofold: first to show that this set of polynomial integrals of motion may be used as the starting point for the construction of a polynomial associative algebra of the type considered in \cite{isaac} and, consequently, for an approach in terms of a deformed oscillator algebra, and second to prove that some rational extensions of Lissajous systems can be constructed via combination of ladder, shift, and supercharge operators and are amenable to a similar analysis.\par
%
%
In Sec.~II, the quantum Lissajous systems considered in \cite{calzada14a, calzada14b} are presented and some of their rational extensions are constructed. The procedure used to derive polynomial integrals of motion is then reviewed. Section~III is devoted to obtain the realizations as deformed oscillator algebra with two approaches, one based on the polynomial associative algebras of the polynomial integrals of motion and the other relying on the generalized Heisenberg algebra generated by the nonpolynomial integrals. In Sec.~IV, we obtain the finite-dimensional unitary representations (unirreps) and the corresponding energy spectrum, which we compare with the physical spectrum. Finally, Sec.~V contains the conclusion.\par
%
%
\section{QUANTUM SUPERINTEGRABLE LISSAJOUS SYSTEMS AND THEIR RATIONAL EXTENSIONS}

\subsection{Definition of the systems}

The Lissajous systems on the sphere \cite{calzada14a, calzada14b} correspond to the Hamiltonians
\begin{equation}
\begin{split}
  H &= - \partial_{\theta}^2 - \cot\theta \partial_{\theta} - \frac{1}{\sin^2 \theta} \partial_{\varphi}^2 -
      \frac{k^2}{\sin^2 \theta} \left(\frac{1/4 - \alpha^2}{\cos^2 k\varphi} 
      + \frac{1/4 - \beta^2}{\sin^2 k\varphi}\right) \\
  &= - \partial_{\theta}^2 - \cot\theta \partial_{\theta} + \frac{k^2}{\sin^2 \theta} \left(- \partial_{\phi}^2
      + \frac{\alpha^2 - 1/4}{\cos^2 \phi} + \frac{\beta^2 - 1/4}{\sin^2 \phi}\right),
\end{split} \label{eq:lissajous}
\end{equation}
where $0 < \theta < \pi$, $k=m/n$ is a rational number, and either one or two terms are present in the potential. In the former case, we assume $\alpha \ge 1$, $\beta = 1/2$, $-\pi/(2k) < \varphi < \pi/(2k)$, and $-\pi/2 < \phi = k\varphi < \pi/2$, whereas in the latter, we suppose $\alpha \ge 1$, $\beta \ge 1$, $0 < \varphi < \pi/(2k)$, and $0 < \phi = k\varphi < \pi/2$. The range of the parameters taken here ensures that the potentials are strongly repulsive at the ends of the interval, the singularities being there of the type $g/\phi^2$ ($g \ge 3/4$) or $g/(\pi/2-\phi)^2$ ($g \ge 3/4$). This means that at each extremity, only one basis solution is quadratically integrable and that the corresponding Hamiltonian is essentially self-adjoint \cite{frank, lathouwers, znojil}.\par
%
%
The Hamiltonian (\ref{eq:lissajous}) is separable in the coordinates $\theta$, $\phi$ and the corresponding eigenvalue equation
\begin{equation}
  H \Psi^{(K)}_{\mu,\nu} (\theta,\phi) = E^K_{\mu} \Psi^{(K)}_{\mu,\nu} (\theta,\phi)  \label{eq:SE}
\end{equation}
has the solutions
\begin{equation}
  \Psi^{(K)}_{\mu,\nu} (\theta,\phi) = \Theta^K_{\mu}(\theta) \Phi_{\nu}(\phi), \qquad E^K_{\mu} =
  (K+\mu)(K+\mu+1), \qquad \mu, \nu = 0, 1, 2, \ldots,  \label{eq:E-H-phi1}
\end{equation}
where
\begin{equation}
\begin{split}
  & \Theta^K_{\mu}(\theta) = N^K_{\mu} \sin^K\theta\, C^{(K+\frac{1}{2})}_{\mu}(-\cos \theta), \\
  & N^K_{\mu} = (-1)^{\mu} 2^K \Gamma\left(K+\frac{1}{2}\right) \left(\frac{\mu! \left(\mu+K+\frac{1}{2}
       \right)}{\pi \Gamma(\mu+2K+1)}\right)^{1/2},
\end{split}  \label{eq:Theta}
\end{equation}
and
\begin{equation}
\begin{split}
  & \Phi_{\nu}(\phi) = N_{\nu} \cos^{\lambda}\phi\, C^{(\lambda)}_{\nu}(\sin\phi), \qquad \lambda = 
      \tfrac{1}{2}(1+2\alpha), \\
  & N_{\nu} = 2^{\lambda-\frac{1}{2}} \Gamma(\lambda) \left(\frac{\nu! (\nu+\lambda)}{\pi \Gamma(\nu
      +2\lambda)}\right)^{1/2}, 
\end{split} \label{eq:Phi-1}
\end{equation}
or
\begin{equation}
\begin{split}
  & \Phi_{\nu}(\phi) = N_{\nu} (\cos\phi)^{\alpha+\frac{1}{2}} (\sin\phi)^{\beta+\frac{1}{2}}
       P^{(\alpha,\beta)}_{\nu}(-\cos 2\phi), \\
  & N_{\nu} = (-1)^{\nu} \left(\frac{2 (\alpha+\beta+1+2\nu) \nu! \Gamma(\alpha+\beta+1+\nu)}
       {\Gamma(\alpha+1+\nu) \Gamma(\beta+1+\nu)}\right)^{1/2}, 
\end{split}  \label{eq:Phi-2}
\end{equation}
in the one- or two-parameter case, respectively. Here $C^{(\lambda)}_{\nu}(x)$ denotes a Gegenbauer polynomial and $P^{(\alpha,\beta)}_{\nu}(x)$ a Jacobi one \cite{abramowitz}. The functions $\Phi_{\nu}(\phi)$, given in (\ref{eq:Phi-1}) or (\ref{eq:Phi-2}), satisfy the equation
\begin{equation}
  H_{\phi} \Phi_{\nu}(\phi) = \epsilon_{\nu}^2 \Phi_{\nu}(\phi), \qquad H_{\phi} = - \partial_{\phi}^2
  + \frac{\alpha^2 - 1/4}{\cos^2\phi}, \qquad \epsilon_{\nu} = \lambda + \nu, \qquad \nu=0, 1, 2, \ldots,
  \label{eq:H_phi-1}
\end{equation}
or 
\begin{equation}
\begin{split}
  & H_{\phi} \Phi_{\nu}(\phi) = \epsilon_{\nu}^2 \Phi_{\nu}(\phi), \qquad H_{\phi} = - \partial_{\phi}^2
     + \frac{\alpha^2 - 1/4}{\cos^2\phi} + \frac{\beta^2 - 1/4}{\sin^2\phi}, \\
  & \epsilon_{\nu} = \alpha + \beta + 1 + 2\nu, \qquad \nu=0, 1, 2, \ldots,
\end{split}  \label{eq:H_phi-2}
\end{equation}
according to which case applies, while $\Theta^K_{\mu}(\theta)$ fulfils
\begin{equation}
\begin{split}
  & H^K_{\theta} \Theta^K_{\mu}(\theta) = E^K_{\mu} \Theta^K_{\mu}(\theta), \qquad H^K_{\theta} =
     - \partial_{\theta}^2 - \cot\theta \partial_{\theta} + \frac{K^2}{\sin^2\theta}, \\
  & K = k \epsilon_{\nu}, \qquad \mu=0, 1, 2, \ldots.  
\end{split} \label{eq:H_theta}
\end{equation}
\par
%
%
To construct a rational extension of $H$, defined in (\ref{eq:lissajous}), is an easy task because $H_{\phi}$ of (\ref{eq:H_phi-2}) is a two-parameter trigonometric P\"oschl-Teller (TPT) Hamiltonian, whose extensions have been thoroughly studied (see, e.g., \cite{cq08, cq09, odake09, odake11, odake13, gomez14}). Here we will restrict ourselves to type I extensions, constructed as one-step isospectral partners of a standard TPT Hamiltonian. At this stage, it is worth observing that the other isospectral extensions, namely type II ones, are related to them because type I and II Jacobi EOP are mirror images of one another, obtained by permuting the roles of $\alpha$ and $\beta$ and changing $\phi$ into $\frac{\pi}{2} - \phi$.\par
%
%
In the case we consider, to get a rational extension of a TPT potential of parameters $\alpha$, $\beta$ in supersymmetric quantum mechanics (SUSYQM) \cite{fernandez}, we have to start from a TPT one of parameters $\alpha+1$, $\beta-1$ and use a supercharge operator
\begin{equation}
  {\cal A} = \partial_{\phi} - \partial_{\phi} \log \chi_{m_1}(\phi), \qquad \chi_{m_1}(\phi) = 
  (\cos\phi)^{-\alpha-\frac{1}{2}} (\sin\phi)^{\beta-\frac{1}{2}} P^{(-\alpha-1,\beta-1)}_{m_1}(-\cos2\phi),
  \label{eq:supercharge}
\end{equation}
with $\alpha>m_1-1$ \cite{footnote}, together with its Hermitian conjugate ${\cal A}^{\dagger}$. Then we arrive at a set of partners on the sphere
\begin{equation}
  H^{(1)} = - \partial_{\theta}^2 - \cot\theta \partial_{\theta} + \frac{k^2}{\sin^2 \theta} 
  \left(- \partial_{\phi}^2 + \frac{(\alpha+3/2)(\alpha+1/2)}{\cos^2 \phi} + \frac{(\beta-1/2)(\beta-3/2)}
  {\sin^2 \phi}\right)
\end{equation}
and    
\begin{equation}
\begin{split}
  H^{(2)} &= - \partial_{\theta}^2 - \cot\theta \partial_{\theta} + \frac{k^2}{\sin^2 \theta} 
     \biggl(- \partial_{\phi}^2 + \frac{\alpha^2 - 1/4}{\cos^2 \phi} + \frac{\beta^2 - 1/4}{\sin^2 \phi} \\
  & \quad - 2 \partial_{\phi}^2 \log P^{(-\alpha-1,\beta-1)}_{m_1}(-\cos2\phi)\biggr), 
\end{split} \label{eq:ext-lissajous}
\end{equation}
where $\alpha>m_1-1$. The additional restrictions on the parameters coming from the behaviour of the set of potentials at the ends of the interval now read $\alpha\ge1$, $\beta\ge2$.\par
%
%
The third Hamiltonian $H$ defined on the sphere we are going to consider here corresponds to $H^{(2)}$ given in (\ref{eq:ext-lissajous}). It still satisfies Eqs.~(\ref{eq:SE})--(\ref{eq:Theta}) and (\ref{eq:H_theta}), but with  different functions $\Phi_{\nu}(\phi)$ resulting from the action of $\cal A$ on the wavefunctions of Eq.~(\ref{eq:Phi-2}) with $(\alpha, \beta) \to (\alpha+1, \beta-1)$. Hence, in this case, Eqs.~(\ref{eq:Phi-2}) and (\ref{eq:H_phi-2}) are replaced by
\begin{equation}
\begin{split}
  & \Phi_{\nu}(\phi) = N_{\nu} \frac{{\cal W}\left(\chi_{m_1}(\phi), (\cos\phi)^{\alpha+\frac{3}{2}} 
      (\sin\phi)^{\beta-\frac{1}{2}} P^{(\alpha+1,\beta-1)}_{\nu}(-\cos 2\phi)\right)}{\chi_{m_1}(\phi)}, \\
  & N_{\nu} = (-1)^{\nu} \left(\frac{(\alpha+\beta+1+2\nu) \nu! \Gamma(\alpha+\beta+1+\nu)}
       {2 (\alpha+\nu-m_1+1) (\beta+\nu+m_1)\Gamma(\alpha+2+\nu) \Gamma(\beta+\nu)}\right)^{1/2},
\end{split} \label{eq:Phi-3}
\end{equation}
and
\begin{equation}
\begin{split}
  & H_{\phi} \Phi_{\nu}(\phi) = \epsilon_{\nu}^2 \Phi_{\nu}(\phi), \\
  & H_{\phi} = - \partial_{\phi}^2 + \frac{\alpha^2 - 1/4}{\cos^2\phi} + \frac{\beta^2 - 1/4}
     {\sin^2\phi} - 2 \partial_{\phi}^2 \log P^{(-\alpha-1,\beta-1)}_{m_1}(-\cos2\phi), \\
  & \epsilon_{\nu} = \alpha + \beta + 1 + 2\nu, \qquad \nu=0, 1, 2, \ldots,
\end{split} \label{eq:H_phi-3} 
\end{equation}
respectively. In Eq.~(\ref{eq:Phi-3}), ${\cal W}(f_1,f_2)$ denotes a Wronskian of $f_1$ and $f_2$.\par
%
%
\subsection{Parameter-dependent integrals of motion}

The procedure used in \cite{calzada14a, calzada14b} to construct integrals of motion for $H$, defined in (\ref{eq:lissajous}), starts by combining  shift operators for $H^K_{\theta}$ with ladder operators for $H_{\phi}$. The Hamiltonian $H^K_{\theta}$ being the same for the three Hamiltonians on the sphere considered here, the same is true for their shift operators. According to \cite{calzada14a, calzada14b}, they are given by
\begin{equation}
  A^+_K = - \partial_{\theta} + (K-1) \cot\theta, \qquad A^-_K = \partial_{\theta} + K \cot\theta. 
  \label{eq:shift}
\end{equation}
\par
%
%
{}For the ladder operators, the situation is more complicated since the three Hamiltonians $H_{\phi}$ are different. For the one- and two-parameter cases, i.e., for (\ref{eq:H_phi-1}) and (\ref{eq:H_phi-2}), they read
\begin{equation}
  B^+_{\nu} = - \cos\phi \partial_{\phi} + (\lambda+\nu) \sin\phi, \qquad B^-_{\nu} = \cos\phi \partial_{\phi} 
  + (\lambda+\nu+1) \sin\phi, \label{eq:ladder-1} 
\end{equation}
and
\begin{equation}
\begin{split}
  B^+_{\nu} &= (\alpha+\beta+2+2\nu) \sin2\phi \partial_{\phi} + (\alpha+\beta+1+2\nu) (\alpha+\beta+2
      +2\nu) \cos2\phi \\
  & \quad - \alpha^2 + \beta^2, \\
  B^-_{\nu} &= - (\alpha+\beta+2\nu) \sin2\phi \partial_{\phi} + (\alpha+\beta+1+2\nu) (\alpha+\beta+2\nu) 
       \cos2\phi - \alpha^2 + \beta^2,
\end{split} \label{eq:ladder-2}
\end{equation}
respectively. As is usual in SUSYQM \cite{fernandez}, those for the rationally-extended partner (\ref{eq:H_phi-3}) can be obtained by combining the ladder operators for the starting Hamiltonian, i.e., the operators (\ref{eq:ladder-2}) with $(\alpha,\beta) \to (\alpha+1,\beta-1)$, with the supercharge operator (\ref{eq:supercharge}) and its Hermitian conjugate, thus yielding
\begin{equation}
\begin{split}
  B^+_{\nu} &= {\cal A} [(\alpha+\beta+2+2\nu) \sin2\phi \partial_{\phi} + (\alpha+\beta+1+2\nu) 
      (\alpha+\beta+2+2\nu) \cos2\phi \\
  & \quad + (\beta-\alpha-2) (\beta+\alpha)] {\cal A}^{\dagger}, \\
  B^-_{\nu} &= {\cal A} [- (\alpha+\beta+2\nu) \sin2\phi \partial_{\phi} + (\alpha+\beta+1+2\nu) 
      (\alpha+\beta+2\nu) \cos2\phi \\
  & \quad + (\beta-\alpha-2) (\beta+\alpha)] {\cal A}^{\dagger}.
\end{split} \label{eq:ladder-3} 
\end{equation}
\par
%
%
On using known properties of Gegenbauer and Jacobi polynomials \cite{abramowitz}, as well as SUSYQM \cite{fernandez}, we have determined the explicit action of the operators (\ref{eq:shift})--(\ref{eq:ladder-3}) on the corresponding wavefunctions. For the operators (\ref{eq:shift}) and (\ref{eq:ladder-1}), for instance, we obtain
\begin{equation}
  A^+_K \Theta^{K-1}_{\mu} = [\mu(\mu+2K-1)]^{1/2} \Theta^K_{\mu-1}, \qquad A^-_K \Theta^K_{\mu} = 
  [(\mu+1)(\mu+2K)]^{1/2} \Theta^{K-1}_{\mu+1},
\end{equation}
and
\begin{equation}
\begin{split}
  & B^+_{\nu} \Phi_{\nu} = \left(\frac{(\nu+1)(\nu+\lambda)(\nu+2\lambda)}{\nu+\lambda+1}\right)^{1/2}
       \Phi_{\nu+1}, \\
  & B^-_{\nu} \Phi_{\nu+1} = \left(\frac{(\nu+1)(\nu+\lambda+1)(\nu+2\lambda)}{\nu+\lambda}\right)^{1/2}
       \Phi_{\nu},
\end{split} \label{eq:ladder-action-1}
\end{equation}
respectively. Results similar to (\ref{eq:ladder-action-1}) for the two-parameter case and the corresponding rational extension are listed in Appendix A.\par
%
%
On combining the shift operators (\ref{eq:shift}) with the ladder ones (\ref{eq:ladder-1}), valid for the one-parameter Hamiltonian, one can form the operators
\begin{equation}
\begin{split}
  X^+_{\mu,\nu} &= A^+_{K+m} \cdots A^+_{K+2} A^+_{K+1} B^+_{\nu+n-1} \cdots B^+_{\nu+1}
      B^+_{\nu}, \\
  X^-_{\mu,\nu} &= A^-_{K-m+1} \cdots A^-_{K-1} A^-_K B^-_{\nu-n} \cdots B^-_{\nu-2} B^-_{\nu-1}. 
\end{split} \label{eq:Xphi-1}
\end{equation}
Whenever an operator $X^+_{\mu,\nu}$ acts on a wavefunction $\Psi^{(K)}_{\mu,\nu}(\theta, \phi)$ of the total Hamiltonian, the product of operators $B^+_{\nu+n-1} \cdots B^+_{\nu+1} B^+_{\nu}$ transforms $\Phi_{\nu}(\phi)$ into $\Phi_{\nu+n}(\phi)$, hence $\epsilon_{\nu}$ into $\epsilon_{\nu+n} = \epsilon_{\nu} + n$, while the product $A^+_{K+m} \cdots A^+_{K+2} A^+_{K+1}$ changes $\Theta^K_{\mu}(\theta)$ into $\Theta^{K+m}_{\mu-m}(\theta)$. Both transformations are compatible because $K = k\epsilon_{\nu} = m \epsilon_{\nu}/n$ becomes $m \epsilon_{\nu+n}/n = m (\epsilon_{\nu}+n)/n = K + m$. In this way and with a similar reasoning for $X^-_{\mu,\nu}$, we arrive at the explicit results
\begin{equation}
\begin{split}
  X^+_{\mu,\nu} \Psi^{(K)}_{\mu,\nu} &= \left(\frac{(\lambda+\nu)(\nu+n)!\Gamma(2\lambda+\nu+n)
      \mu!\Gamma(\mu+2K+m+1)}{(\lambda+\nu+n)\nu!\Gamma(2\lambda+\nu)(\mu-m)!\Gamma(\mu+2K+1)}
      \right)^{1/2} \Psi^{(K+m)}_{\mu-m,\nu+n}, \\ 
  X^-_{\mu,\nu} \Psi^{(K)}_{\mu,\nu} &= \left(\frac{(\lambda+\nu)\nu!\Gamma(2\lambda+\nu)(\mu+m)!
      \Gamma(\mu+2K+1)}{(\lambda+\nu-n)(\nu-n)!\Gamma(2\lambda+\nu-n)\mu!\Gamma(\mu+2K-m+1)}
      \right)^{1/2} \\ 
  & \quad \times\Psi^{(K-m)}_{\mu+m,\nu-n},  
\end{split} \label{eq:X-action-1}
\end{equation}
and we observe that the wavefunctions on both sides of these relations correspond to the same energy $E^K_{\mu} = E^{K\pm m}_{\mu\mp m}$. The operators $X^{\pm}_{\mu,\nu}$ therefore commute with $H$ when acting on $\Psi^{(K)}_{\mu,\nu}$. One may also note that the successive application of two operators of type $X^{\pm}_{\mu,\nu}$ leads to a multiple of the initial eigenfunction,
\begin{equation}
\begin{split}
  X^+_{\mu+m,\nu-n} X^-_{\mu,\nu} \Psi^{(K)}_{\mu,\nu} &= \frac{\nu!\Gamma(2\lambda+\nu)(\mu+m)!
       \Gamma(\mu+2K+1)}{(\nu-n)!\Gamma(2\lambda+\nu-n)\mu!\Gamma(\mu+2K-m+1)} 
       \Psi^{(K)}_{\mu,\nu}, \\
  X^-_{\mu-m,\nu+n} X^+_{\mu,\nu} \Psi^{(K)}_{\mu,\nu} &= \frac{(\nu+n)!\Gamma(2\lambda+\nu+n)\mu!
       \Gamma(\mu+2K+m+1)}{\nu!\Gamma(2\lambda+\nu)(\mu-m)!\Gamma(\mu+2K+1)} 
       \Psi^{(K)}_{\mu,\nu}.
\end{split} \label{eq:X-product-1}
\end{equation}
\par
%
%
{}For the two-parameter Hamiltonian (\ref{eq:lissajous}) and its rational extension (\ref{eq:ext-lissajous}), one considers instead the operators
\begin{equation}
\begin{split}
  X^+_{\mu,\nu} &= A^+_{K+2m} \cdots A^+_{K+2} A^+_{K+1} B^+_{\nu+n-1} \cdots B^+_{\nu+1}
      B^+_{\nu}, \\
  X^-_{\mu,\nu} &= A^-_{K-2m+1} \cdots A^-_{K-1} A^-_K B^-_{\nu-n+1} \cdots B^-_{\nu-1} B^-_{\nu}, 
\end{split} \label{eq:XSE-phi-3}
\end{equation}
which change $\Psi^{(K)}_{\mu,\nu}(\theta, \phi)$ into $\Psi^{(K\pm2m)}_{\mu\mp2m,\nu\pm n}(\theta, \phi)$ with the same energy $E^K_{\mu} = E^{K\pm 2m}_{\mu\mp 2m}$. Their explicit action can be found in Appendix A.\par
%
%
\subsection{Parameter-independent (nonpolynomial) integrals of motion}

Since the operators $X^{\pm}_{\mu\pm m,\nu\mp n} X^{\mp}_{\mu,\nu}$ of Eq.~(\ref{eq:X-product-1}) leave $\Psi^{(K)}_{\mu,\nu}(\theta,\phi)$ essentially unchanged, they have the same effect on the latter as some functions of $H$ and $H_{\phi}$. As suggested by Calzada et al.~\cite{calzada14a, calzada14b}, we therefore replace them by the parameter-independent products
\begin{equation}
\begin{split}
  X^+ X^- &= \prod_{r=1}^n \left[\left(\sqrt{H_{\phi}}-r\right)\left(\sqrt{H_{\phi}}-r+1\right) - \alpha^2
      + \tfrac{1}{4}\right] \\
  & \quad \times \prod_{p=1}^m \left[H - \left(k\sqrt{H_{\phi}}-p\right)\left(k\sqrt{H_{\phi}}-p
      +1\right)\right], \\
  X^- X^+ &= \prod_{r=1}^n \left[\left(\sqrt{H_{\phi}}+r\right)\left(\sqrt{H_{\phi}}+r-1\right) - \alpha^2
      + \tfrac{1}{4}\right] \\
 & \quad \times \prod_{p=1}^m \left[H - \left(k\sqrt{H_{\phi}}+p\right)\left(k\sqrt{H_{\phi}}+p
      -1\right)\right]. 
\end{split} \label{eq:X-product-1-bis}
\end{equation}
By proceeding similarly for the products of operators (\ref{eq:X-product-2}) and (\ref{eq:X-product-3}), valid for the two-parameter Hamiltonian and its rational extension, we get
\begin{equation}
\begin{split}
  X^+ X^- &= \prod_{r=1}^n \Bigl\{\left[\left(\sqrt{H_{\phi}}-2r\right)\left(\sqrt{H_{\phi}}-2r+2\right) - 
      (\alpha+\beta+1)(\alpha+\beta-1)\right] \\
  & \quad \times \left[\left(\sqrt{H_{\phi}}-2r\right)\left(\sqrt{H_{\phi}}-2r+2\right) - (\alpha-
      \beta+1)(\alpha-\beta-1)\right]\Bigr\} \\
  & \quad \times \prod_{p=1}^{2m} \left[H - \left(k\sqrt{H_{\phi}}-p\right)\left(k\sqrt{H_{\phi}}-p
      +1\right)\right], \\
  X^- X^+ &= \prod_{r=1}^n \Bigl\{\left[\left(\sqrt{H_{\phi}}+2r\right)\left(\sqrt{H_{\phi}}+2r-2\right) -
      (\alpha+\beta+1)(\alpha+\beta-1)\right] \\
  & \quad \times \left[\left(\sqrt{H_{\phi}}+2r\right)\left(\sqrt{H_{\phi}}+2r-2\right) - (\alpha-
      \beta+1)(\alpha-\beta-1)\right]\Bigr\} \\
  & \quad \times \prod_{p=1}^{2m} \left[H - \left(k\sqrt{H_{\phi}}+p\right)\left(k\sqrt{H_{\phi}}+p
      -1\right)\right], 
\end{split} \label{eq:X-product-2-bis}
\end{equation}
and
\begin{equation}
\begin{split}
  &X^+ X^- \\
  &= \prod_{q=1}^n \Bigl\{\left[\left(\sqrt{H_{\phi}}-2q-1\right)\left(\sqrt{H_{\phi}}-2q+1\right)
      - (\alpha-\beta-2m_1)(\alpha-\beta-2m_1+2)\right] \\ 
  & \quad \times \left[\left(\sqrt{H_{\phi}}-2q+1\right)\left(\sqrt{H_{\phi}}-2q+3\right)
      - (\alpha-\beta-2m_1)(\alpha-\beta-2m_1+2)\right]\Bigr\} \\  
  & \quad \times \prod_{r=1}^n \Bigl\{\left[\left(\sqrt{H_{\phi}}-2r\right)\left(\sqrt{H_{\phi}}-2r+2\right) - 
      (\alpha+\beta+1)(\alpha+\beta-1)\right] \\
  & \quad \times \left[\left(\sqrt{H_{\phi}}-2r\right)\left(\sqrt{H_{\phi}}-2r+2\right) - (\alpha-
      \beta+3)(\alpha-\beta+1)\right]\Bigr\} \\
  & \quad \times \prod_{p=1}^{2m} \left[H - \left(k\sqrt{H_{\phi}}-p\right)\left(k\sqrt{H_{\phi}}-p
      +1\right)\right], \\
  &X^- X^+ \\
  &= \prod_{q=1}^n \Bigl\{\left[\left(\sqrt{H_{\phi}}+2q+1\right)\left(\sqrt{H_{\phi}}+2q-1\right)
      - (\alpha-\beta-2m_1)(\alpha-\beta-2m_1+2)\right] \\ 
  & \quad \times \left[\left(\sqrt{H_{\phi}}+2q-1\right)\left(\sqrt{H_{\phi}}+2q-3\right)
      - (\alpha-\beta-2m_1)(\alpha-\beta-2m_1+2)\right]\Bigr\} \\  
  & \quad \times \prod_{r=1}^n \Bigl\{\left[\left(\sqrt{H_{\phi}}+2r\right)\left(\sqrt{H_{\phi}}+2r-2\right) -
      (\alpha+\beta+1)(\alpha+\beta-1)\right] \\
  & \quad \times \left[\left(\sqrt{H_{\phi}}+2r\right)\left(\sqrt{H_{\phi}}+2r-2\right) - (\alpha-
      \beta+3)(\alpha-\beta+1)\right]\Bigr\} \\
  & \quad \times \prod_{p=1}^{2m} \left[H - \left(k\sqrt{H_{\phi}}+p\right)\left(k\sqrt{H_{\phi}}+p
      -1\right)\right], 
\end{split} \label{eq:X-product-3-bis}
\end{equation}
respectively.\par
%
%
In all three cases, we have a set of four (formal) integrals of motion $(H, H_{\phi}, X^{\pm})$, which, as shown in Eqs.~(\ref{eq:X-product-1-bis})--(\ref{eq:X-product-3-bis}), are algebraically dependent and nonpolynomial since the square root operator $\sqrt{H_{\phi}}$ appears everywhere. We may actually decompose $X^+ X^-$ and $X^- X^+$ as
\begin{equation}
  X^+X^- = P_1(H,H_{\phi}) - P_2(H,H_{\phi}) \sqrt{H_{\phi}}, \qquad X^-X^+ = P_1(H,H_{\phi}) +
  P_2(H,H_{\phi}) \sqrt{H_{\phi}}, \label{eq:decomposition}
\end{equation}
where $P_1(H,H_{\phi})$ and $P_2(H,H_{\phi})$ are polynomials in $H$ and $H_{\phi}$ of respective degrees $(n+m,n+m-1)$, $(2n+2m,2n+2m-1)$, and $(4n+2m,4n+2m-1)$ in the three cases (\ref{eq:X-product-1-bis}), (\ref{eq:X-product-2-bis}), and (\ref{eq:X-product-3-bis}).\par
%
%
In the one-parameter case, the action of $H_{\phi}$ and $X^{\pm}_{\mu,\nu}$ on $\Psi^{(K)}_{\mu,\nu}(\theta,\phi)$, given in Eqs.~(\ref{eq:H_phi-1}) and (\ref{eq:X-action-1}), directly leads to the (formal) commutator
\begin{equation}
  \left[\sqrt{H_{\phi}}, X^{\pm}\right] = \pm n X^{\pm}, \label{eq:alg-1}
\end{equation}
which implies
\begin{equation}
  [H_{\phi}, X^{\pm}] = X^{\pm} \left(\pm 2n \sqrt{H_{\phi}} + n^2\right). \label{eq:alg-2}
\end{equation}
The latter may be completed by a commutator and a restriction relation, directly deriving from (\ref{eq:decomposition}),
\begin{equation}
  [X^+, X^-] = -2 P_2(H,H_{\phi}) \sqrt{H_{\phi}}, \qquad X^+X^- + X^-X^+ = 2 P_1(H,H_{\phi}).
  \label{eq:alg-3}
\end{equation}
\par
%
%
In the two-parameter case and its rational extension, Eq.~(\ref{eq:alg-3}) remains valid, but Eqs.~(\ref{eq:alg-1}) and (\ref{eq:alg-2}) are replaced by
\begin{equation}
  \left[\sqrt{H_{\phi}}, X^{\pm}\right] = \pm 2n X^{\pm}, \qquad [H_{\phi}, X^{\pm}] = X^{\pm} \left(\pm 4n
  \sqrt{H_{\phi}} + 4n^2\right),
\end{equation}
where $2n$ has been substituted for $n$.\par
%
%
\subsection{Polynomial integrals of motion}

{}From the nonpolynomial (formal) integrals of motion $X^{\pm}$, it is now possible to build some polynomial ones by decomposition. Let us indeed set
\begin{equation}
  X^+ = {\cal O}\sqrt{H_{\phi}} + {\cal E}, \qquad X^- = \varepsilon \left(-{\cal O}\sqrt{H_{\phi}} + {\cal E}
  \right), \label{eq:decomposition-bis}
\end{equation}
where $\cal O$ and $\cal E$ may be written as polynomial differential operators in $\partial_{\theta}$, $\partial_{\phi}$ of respective orders $(n+m-1,n+m)$, $(2n+2m-1,2n+2m)$, and $(4n+2m-1,4n+2m)$ in the three cases we consider. Here $\varepsilon$ is some sign defined by $\varepsilon = (-1)^{n+m}$, $\varepsilon = (-1)^{2n+2m} = +1$, and $\varepsilon = (-1)^{4n+2m} = +1$, respectively. Since $X^{\pm}$ are integrals of motion, the same is true for $\cal O$ and $\cal E$. We therefore arrive at a set $(H, H_{\phi}, {\cal O}, {\cal E})$ of four (formal) polynomial, albeit non algebraically independent, integrals of motion.\par
%
%
Inserting the definitions (\ref{eq:decomposition-bis}) in the relations (\ref{eq:alg-1})--(\ref{eq:alg-3}) leads to the commutation relations
\begin{equation}
  \left[\sqrt{H_{\phi}}, {\cal O}\right] \sqrt{H_{\phi}} = n {\cal E}, \qquad \left[\sqrt{H_{\phi}}, {\cal E}\right]
  = n {\cal O} \sqrt{H_{\phi}},
\end{equation}
\begin{equation}
  [H_{\phi}, {\cal O}] = n^2 {\cal O} + 2n {\cal E}, \quad [H_{\phi}, {\cal E}] = 2n {\cal O}H_{\phi} + n^2
  {\cal E}, \quad [{\cal O}, {\cal E}] = -n {\cal O}^2 - \varepsilon P_2(H,H_{\phi}), \label{eq:alg-1-bis}
\end{equation}
and the restriction relation
\begin{equation}
  - {\cal O}^2 H_{\phi} + {\cal E}^2 - n {\cal O}{\cal E} = \varepsilon P_1(H,H_{\phi}). \label{eq:alg-2-bis}
\end{equation}
As before, the two-parameter case and its rational extension are simply dealt with by replacing $n$ by $2n$ and by taking care of the appropriate definition of $\varepsilon$ in these equations.\par
%
%
Since the Hermitian properties of $X^{\pm}$ with respect to the inner product corresponding to the measure $\sin\theta d\theta d\phi$ are not known a priori, the same is true for those of $\cal O$ and $\cal E$. Calzada et al.~\cite{calzada14a, calzada14b} have devised a procedure to determine them, which uses both the order of the differential operators and the fact that their Hermitian properties must preserve the commutation relations (\ref{eq:alg-1-bis}), as well as the restriction relation (\ref{eq:alg-2-bis}). The results read
\begin{equation}
  {\cal O}^{\dagger} = - \varepsilon {\cal O}, \qquad {\cal E}^{\dagger} = \varepsilon ({\cal E}+n{\cal O}).
  \label{eq:hermite}
\end{equation}
It then follows that the new set of polynomial integrals of motion $(H,H_{\phi},{\cal O},{\cal E}')$, with
\begin{equation}
  {\cal E}' = {\cal E} + \tfrac{1}{2}n {\cal O},  \label{eq:Ep}
\end{equation}
only contains Hermitian/anti-Hermitian operators since
\begin{equation}
  {\cal O}^{\dagger} = - \varepsilon {\cal O}, \qquad {\cal E}^{\prime\dagger} = \varepsilon {\cal E}'.
\end{equation}
In terms of these new operators, Eqs.~(\ref{eq:alg-1-bis}) and (\ref{eq:alg-2-bis}) become
\begin{equation}
  [H_{\phi}, {\cal O}] = 2n {\cal E}', \quad [H_{\phi}, {\cal E}'] = n \{H_{\phi}, {\cal O}\} - \tfrac{1}{2} n^3
  {\cal O}, \quad [{\cal O}, {\cal E}'] = -n {\cal O}^2 - \varepsilon P_2(H,H_{\phi}), \label{eq:alg-ter0}
\end{equation}
and
\begin{equation}
  - {\cal O} H_{\phi} {\cal O} + {\cal E}^{\prime2} + \tfrac{1}{4} n^2 {\cal O}^2 = \varepsilon 
  \left[P_1(H,H_{\phi}) + \tfrac{1}{2} n P_2(H,H_{\phi})\right], \label{eq:alg-ter}
\end{equation}
respectively. Here $\{A,B\}$ denotes the anticommutator of $A$ and $B$. Equations (\ref{eq:hermite})--(\ref{eq:alg-ter}) remain valid for the two-parameter case and its rational extension after the appropriate changes in $n$ and $\varepsilon$ have been carried out.\par
%
%
\section{REALIZATION AS DEFORMED OSCILLATOR ALGEBRAS}
\setcounter{equation}{0}
 
In this Section, we will obtain the realizations as deformed oscillator algebra for the three models corresponding to Eqs.~(\ref{eq:H_phi-1}), (\ref{eq:H_phi-2}), and (\ref{eq:H_phi-3}) using two approaches. One of these methods is based on the finitely generated polynomial algebra that is constructed from the polynomial integrals of motion ${\cal E}'$ and ${\cal O}$, given by Eqs.~(\ref{eq:decomposition-bis}) and (\ref{eq:Ep}).  The other approach uses the nonpolynomial integrals $X^{+}$ and $X^{-}$, obtained from Eqs.~(\ref{eq:Xphi-1}), (\ref{eq:XSE-phi-3}), and satisfying with $H_{\phi}$ a generalized Heisenberg algebra (GHA).\par
%
%
\subsection{Daskaloyannis approach and polynomial algebras}

The most general polynomial algebra for a 2D superintegrable system with an integral of second order and another one of arbitrary order $M$ was studied in Ref.~\cite{isaac}. The set of constraints on the structure constants, obtained by imposing the Jacobi identity and using various commutator and anticommutator identities, was presented. In addition, a way was devised to construct the Casimir operator and to obtain the realizations as deformed oscillator algebras.\par
%
%
Among the general class of finitely generated polynomial algebras with only three generators studied, let us take a particular case relevant to this paper and the quantum models considered:
\begin{subequations}
\begin{equation}
  [A,B]=C,  \label{eq:pola1}
\end{equation}
\begin{equation}
  [A,C]=\alpha_{1} A+\alpha_{2}A^{2}+\delta B+\epsilon +\beta \{A,B\}, \label{eq:pola2}
\end{equation}
\begin{equation}
  [B,C]=\sum_{i=1}^M\lambda_{i}A^{i}-\beta B^{2}-\alpha_{1} B -\alpha_{2}\{A,B\} + \zeta, \label{eq:pola3}
\end{equation}
\end{subequations}
where $A$, $B$ are two Hermitian operators (so that $C$ is anti-Hermitian) and $\alpha_{1}$, $\alpha_{2}$, $\delta$, $\epsilon$, $\beta$, $\lambda_{i}$, $\zeta$ in the commutation relations (\ref{eq:pola1}), (\ref{eq:pola2}), and (\ref{eq:pola3}) are the structure constants. Let us point out that they are not simply constants as they can be polynomials of the Hamiltonian, which is a central element of the polynomial associative algebra. The maximal order of these polynomials in terms of $H$ can be deduced from the order as differential operator of the commutator on the left-hand side.\par
%
%
This algebra has a Casimir operator, which can be expressed in terms of $A$, $B$, and $C$ as
\begin{equation}
\begin{split}
  K &= C^2 - (\alpha_1-\beta\alpha_2) \{A,B\} - \alpha_2 \{A^2,B\} - \beta \{A,B^2\} + \sum_{i=1}^{M+1}
      k_i A^i - (2\epsilon-\alpha_1\beta) B \\
  & \quad {}+ (\beta^2-\delta) B^2,  
\end{split}\label{eq:casimir}
\end{equation}
where $k_i$, $i=1$, 2, \ldots, $M+1$, are some parameters that can be determined in principle to any given order by solving a set of constraints. In the case of integrals of arbitrary order $M$, no explicit solutions are known to these constraints.\par
%
%
Moreover, this class of algebras admits a realization as deformed oscillator algebra of the form
\begin{equation}
  A=A(N), \quad B=B_0(N)+b^{\dagger}\rho(N)+\rho(N)b,  \label{eq:real}
\end{equation}
where the generators of the deformed oscillator algebra $\{b,b^{\dagger},N,1\}$ satisfy
\begin{equation}
  [N,b^{\dagger}]=b^{\dagger},\quad [N,b]=-b, \quad b^{\dagger}b=\Phi(N),\quad bb^{\dagger}=\Phi(N+1),    
  \label{eq:doa}
\end{equation}
and $\Phi(N)$ is referred to as a structure function \cite{daska}. The functions $A(N)$ and $B_{0}(N)$ in Eq.~(\ref{eq:real}) take the following form \cite{isaac}
\begin{equation}
  A(N) =\frac{\beta}{2}\left[ \left((N+u)^{2}-\frac{1}{4}\right)-\frac{\delta}{\beta^{2}}\right],
\end{equation}
\begin{equation}
\begin{split}
  B_0(N) &=-\frac{\alpha_{2}}{4} \left((N+u)^{2}-\frac{1}{4}\right)+\left(\frac{-\beta \alpha_{1}+
      \alpha_{2}\delta}{2\beta^{2}}\right) + \epsilon \\
  & \quad{} -\left(\frac{-2\beta \alpha_{1} \delta +\alpha_{2} \delta^{2}}{4\beta^{4}}\right) 
      \frac{1}{\left((N+u)^{2}-\frac{1}{4}\right)},
\end{split}
\end{equation}
where $u$ is some representation-dependent constant. Provided one is able to get some supplementary information on the Casimir operator, two additional equations linear in $\Phi(N)$ and $\Phi(N+1)$,
\begin{equation}
\begin{split}
  & -2 \Phi(N) \rho(N-1)^{2} \Delta A(N-1) +2 \Phi(N+1) \rho(N)^{2} \Delta A(N) \\ 
  &= \zeta +\sum_{i=1}^{M} \lambda_{i} A(N)^{i}-\alpha_{1} B_0(N) -2 \alpha_2 B_0(N)A(N) \\
  &\quad -\beta [B_0(N)^2 + \rho(N-1)^2 \Phi(N) +\rho(N)^2 \Phi(N+1)],
\end{split}  \label{eq:const1}
\end{equation}
\begin{equation}
\begin{split}
  K &= - [\Delta A(N-1)]^2 \rho(N-1)^2 \Phi(N) - [\Delta A(N)]^2 \rho(N)^2 \Phi(N+1) \\
  &\quad - 2\alpha_1A(N)B_0(N) -2\alpha_2A(N)^2B_0(N) +2\beta\alpha_2A(N)B_0(N) \\ 
  &\quad - 2\beta A(N)B_0(N)^2 - 2\beta A(N)\rho(N-1)^2\Phi(N) -2\beta A(N)\rho(N)^2\Phi(N+1) \\
  &\quad  + \sum_{i=1}^{M+1}k_i A(N)^i - (2\epsilon-\beta\alpha_1)B_0(N) \\
  &\quad  +(\beta^2-\delta)\left[B_0(N)^2 + \rho(N-1)^2\Phi(N) + \rho(N)^2\Phi(N+1)\right],  
\end{split} \label{eq:const2}
\end{equation}
allow to obtain the structure function $\Phi(N)$. Here $\Delta A(N) = A(N+1) - A(N)$.\par
%
%
{}For low order integrals of motion, an expression of $K$ as a differential operator can be determined from that of $A$, $B$, $C$, and the Hamiltonian $H$. From this, $K$ can be rewritten as a polynomial of $H$ only. This turns out, however, to be an involving task even for low values of $M \le 4$ \cite{daska,marquette09,marquette13a}. In recent papers \cite{marquette10,marquette14}, it has been demonstrated, in the case of 2D Hamiltonians with separation of variables in Cartesian coordinates and for which the integrals are generated using combinations of ladder operators, that it is possible to overcome this difficulty in the application of this algebraic method and to get the structure function by exploiting the underlying structure of the polynomial algebra. We will show here how we can also avoid such calculations concerning the Casimir operator for our models with separation in spherical variables by taking advantage of the existence of an algebraic relation in addition to the polynomial algebra commutation relations and the Casimir operator. \par
%
%
Taking the algebra defined by Eq.~(\ref{eq:alg-ter0}), valid for the one-parameter Lissajous system, and setting $A=H_{\phi}$, $B= \eta {\cal O}$, and $C=2n\eta {\cal E}'$ with $\eta = {\rm i}^{m+n-1}$ (so that $A^{\dagger}=A$, $B^{\dagger}=B$, and $C^{\dagger}=-C$), we obtain the standard form (\ref{eq:pola1}), (\ref{eq:pola2}), and (\ref{eq:pola3}), 
\begin{subequations}
\begin{equation}
  [A,B]=C,     \label{eq:polal1-1p}
\end{equation}
\begin{equation}
  [A,C]=  2n^{2}\{A,B\}-n^{4} B, \label{eq:polal2-1p}
\end{equation}
\begin{equation}
  [B,C]=-2n^{2} B^{2} +2 n P_{2}(H,A), \label{eq:polal3-1p}
\end{equation}
\end{subequations}
with the polynomial identity given by Eq.~(\ref{eq:alg-ter}) in terms of the generators $A$, $B$, and $C$ as
\begin{equation}
  C^{2}-2n^{2}\{A,B^{2}\}+5n^{4}B^{2}=-4n^{2}\left(P_{1}-\frac{n}{2}P_{2}\right).  \label{eq:polal4-1p}
\end{equation}
\par
%
%
This is one of the key steps in extending the Daskaloyannis approach for these Hamiltonians, the products $X^{+}X^{-}$ and $X^{-}X^{+}$ providing in fact two relations, the last  commutation relation of the polynomial algebra (\ref{eq:polal3-1p}) and, moreover, this algebraic constraint (\ref{eq:polal4-1p}) that plays an important role in the algebraic derivation. From Eq.~(\ref{eq:casimir}), the Casimir operator takes the form
\begin{equation}
  K=C^{2}-2n^{2}\{A,B^{2}\}+5 n^{4} B^{2} +\sum_{i=1}^{M+1} k_{i} A^{i},
\end{equation}
where the $k_{i}$'s satisfy a set of constraints, as explained above. However, we do not need to solve the latter because using Eq.~(\ref{eq:polal4-1p}), the Casimir operator can be rewritten as
\begin{equation}
  K= -4 n^{2} \left(P_{1}-\frac{n}{2}P_{2}\right) + \sum_{i=1}^{M+1} k_{i} A^{i}.
\end{equation}
Inserting then this alternative form in the left-hand side of Eq.~(\ref{eq:const2}), the system of equations (\ref{eq:const1}) and (\ref{eq:const2}) for the structure function can be simplified and only involves the central element $H$ and the generator $A$, which is directly connected to the number operator. The results read
\begin{equation}
A(N)=n^{2}(N+u)^{2},\quad B_{0}(N)=0,
\end{equation}
\begin{equation}
  \rho^{2}(N)=[4 n^{2}(N+u)(N+u+1)]^{-1},
\end{equation}
\begin{equation}
  \Phi(N)=P_{1}\bigl(H,A(N)\bigr)-n(N+u)P_{2}\bigl(H,A(N)\bigr).
\end{equation}
\par
%
%
In the case of the two-parameter Lissajous system and its rational extensions related to Jacobi EOP of type I (or II), all relations remain valid by replacing $n$ by $2n$ and setting $\eta={\rm i}$. The fact that the structure function can be obtained in this way is highly non trivial and is a consequence of the structure of the algebra generated by $H_{\phi}$, $\cal O$, and ${\cal E}'$. Let us also note that we did not need to obtain explicitly the structure constants of the polynomial algebra, the parameters of the Casimir operator, nor, more importantly, the Casimir operator expressed in terms of the Hamiltonian only.\par
%
%
\subsection{Realization as deformed oscillator algebras from generalized Heisenberg ones}

{}For these models, instead of the polynomial integrals $\cal O$ and ${\cal E}'$, let us consider the nonpolynomial ones $X^+$ and $X^-$. The algebras generated by $\{H_{\phi},X^{+},X^{-}\}$ and defined by Eqs.~(\ref{eq:alg-1}), (\ref{eq:alg-2}), and (\ref{eq:alg-3}), are in fact generalized Heisenberg algebras (GHA) \cite{delbecq93,eleonsky96,quesne99,curado2001,dong07,curado08}, belonging to the following class
of algebraic structures
\begin{equation}
H_{\phi}X^{+}=X^{+}f(H_{\phi}),
\end{equation}
\begin{equation}
X^{-}H_{\phi}=f(H_{\phi})X^{-},
\end{equation}
\begin{equation}
[X^{-},X^{+}]=g(H_{\phi}),
\end{equation}
where $f(z)$ and $g(z)$ are not simply polynomials. Such algebras were observed in various contexts and, in particular, in regard of 1D quantum systems, such as the infinite well as well as the Morse and P\"oschl-Teller potentials. To find such algebraic structures in the context of superintegrable systems and their integrals is highly interesting. \par
%
%
Here, the generalized Heisenberg algebra has a very specific structure. The function $f(H_{\phi})$ involves only square roots of $H_{\phi}$ and not only the commutator $[X^{-},X^{+}]$ is known, but also the products $X^{-}X^{+}$ and $X^{+}X^{-}$ themselves have been calculated explicitly in a convenient product form. It can be shown that the GHA of the one-parameter Lissajous system on the sphere can be put in the form of a deformed oscillator algebra by defining the number operator $N$ through the equation $\sqrt{H_{\phi}}= (N+u)n$, where $u$ is some representation-dependent constant, and by taking $b=X^{-}$ and $b^{\dagger}=X^{+}$. The structure function coincides with the one obtained using the polynomial algebra approach with a Casimir operator and is explicitly given by
\begin{equation}
\begin{split}
  \Phi(N,H,u)&=b^{\dagger}b=X^{+}X^{-}=P_{1}(H,H_{\phi})-P_{2}(H,H_{\phi})\sqrt{H_{\phi}}  \\
  &=P_{1}(H,H_{\phi})-n(N+u)P_{2}(H,H_{\phi}). 
\end{split}
\end{equation}
\par
%
%
In a similar way, for the two-parameter Lissajous system and its rational extensions, we take simply $\sqrt{H_{\phi}}= (N+u)2n$. \par
%
%
Let us now present at this stage, the expression for the structure function of the one-parameter Lissajous ($\Phi^{(1)}(N,H,u)$), the two-parameter Lissajous ($\Phi^{(2)}(N,H,u)$) and the rationally extended two-parameter Lissajous ($\Phi^{(E2)}(N,H,u)$):
\begin{equation}
\begin{split}
  &\Phi^{(1)}(N,H,u) = \prod_{p=1}^{m} \{H-[m(N+u)-p][m(N+u)-p+1]\} \\
  &\quad \times \prod_{r=1}^{n} \left\{ [n(N+u)-r][n(N+u)-r+1]-\alpha^{2}+\frac{1}{4}\right\},    
\end{split}  \label{eq:strut-1pl}
\end{equation}
\begin{equation}
\begin{split}
  &\Phi^{(2)}(N,H,u) = \prod_{p=1}^{2m} \{H-[2m(N+u)-p][2m(N+u)-p+1]\}  \\
  &\quad \times \prod_{r=1}^{n} \{[2n(N+u)-2r][2n(N+u)-2r+2]-(\alpha+\beta+1)(\alpha+\beta-1)\}  \\
  &\quad \times \prod_{r=1}^n \{[2n (N+u)-2r][2n (N+u)-2r +2]-(\alpha-\beta+1)(\alpha-\beta-1)\}, 
\end{split} \label{eq:strut-2pl}
\end{equation}
\begin{equation}
\begin{split}
  &\Phi^{(E2)}(N,H,u) = \prod_{p=1}^{2m} \{H-[2m (N+u)-p][2m (N+u)-p+1]\}  \\
  &\quad \times \prod_{q=1}^{n} \{[ 2n (N+u)-2q-1][2n (N+u)-2q+1] \\
  &\quad \quad \quad -(\alpha-\beta-2m_{1})(\alpha-\beta-2m_{1}+2)\} \\
  &\quad \times \prod_{q=1}^n \{[2n ( N+u) -2q +1][2n (N +u)-2q +3] \\
  &\quad \quad \quad -(\alpha -\beta -2m_{1})(\alpha-\beta-2m_1+2)\} \\
  &\quad \times \prod_{r=1}^n \{[2n(N+u)-2r][2n(N+u)-2r+2] - (\alpha+\beta+1)(\alpha+\beta-1)\} \\
  &\quad \times \prod_{r=1}^n \{[2n(N+u)-2r][2n(N+u)-2r+2] - (\alpha-\beta+3)(\alpha-\beta+1)\}.
\end{split}\label{eq:strut-e2pl}
\end{equation}
\par
%
%
\section{ANALYSIS OF STRUCTURE FUNCTION AND SPECTRUM}
\setcounter{equation}{0}

In this Section, we will use the expressions obtained in two different ways for the structure functions (\ref{eq:strut-1pl}), (\ref{eq:strut-2pl}), and (\ref{eq:strut-e2pl}) and present an algebraic derivation of the spectrum of the models using finite-dimensional unirreps of the deformed oscillator algebra.\par
%
%
The unirreps can be obtained by introducing an energy-dependent Fock space of dimension $\bar{p}+1$, defined by the action of the Hamiltonian $H$, the number operator $N$, and the creation and annihilation operators $b^{\dagger}$, $b$. By acting iteratively on a state with a given energy, these operators $b^{\dagger}$ and $b$ allow to reach every state in the multiplet it belongs. Let us mention that the underlying structure of these unirreps is not always related in a straightforward manner to the physical states of the model, but can be revealed by using a detailed analysis \cite{marquette14}.\par
%
%
 We consider $H|E,n\rangle=E|E,n\rangle$, $N|E,n\rangle=n|E,n\rangle$, and $b|E,0\rangle=0$. The action of the operators $b^{\dagger}$ and $b$ is given by
\begin{equation}
\begin{split}
  & b^{\dagger}|E, n\rangle=\sqrt{\Phi(n+1,E,u)}|E,n+1\rangle, \\
  & b|E, n\rangle=\sqrt{\Phi(n,E,u)}|E,n-1\rangle.
\end{split}
\end{equation}
We see the important role of the structure function in this construction of the Fock space. To obtain the unirreps we further impose the following constraints
\begin{equation}
  \Phi(\bar{p}+1,E,u)=0, \quad \Phi(0,E,u)=0,\quad \Phi(n,E,u)>0 \quad n=1, 2, \ldots, \bar{p}.    
  \label{eq:const123}
\end{equation}
The energy $E$ and the constant $u$ can be obtained from this set of constraints, which are algebraic
equations. The dimension of the finite-dimensional unirreps is given by $\bar{p}+1$. The nonlinearity of the structure function allows the existence of patterns of unirreps  describing a more complicated spectrum and the fact that levels associated with a given energy can be organized in several multiplets of different length. It also enables to obtain different types of solutions that correspond to equivalent ways to enumerate the levels.\par
%
%
\subsection{One-parameter Lissajous system}

The structure function given by Eq.~(\ref{eq:strut-1pl}) can be factorized in the following way:
\begin{equation}
\begin{split}
  & \Phi(x,E,u) = (-1)^m m^{2m} n^{2n} \prod_{r=1}^n \left[\left(x+u-\frac{2r-1-2\alpha}{2n}\right) 
      \left(x+u-\frac{2r-1+2\alpha}{2n}\right)\right] \\
  &\quad \times \prod_{p=1}^m \left[\left(x+u-\frac{2p-1+\sqrt{1+4E}}{2m}\right) 
      \left(x+u-\frac{2p-1-\sqrt{1+4E}}{2m}\right)\right].
\end{split}
\end{equation}
\par
%
%
On using the constraints (\ref{eq:const123}), two equivalent solutions are obtained
\begin{align}
  & u_{1}=\frac{2\tilde{r}-1+2\alpha}{2n}, \quad \tilde{r} \in \{1, 2, \ldots, n\},   \\
  & u_{2}=\frac{2\tilde{p}-1-\sqrt{1+4E}}{2m}, \quad \tilde{p} \in \{1, 2, \ldots, m\}, 
\end{align}
with the corresponding energy spectrum   
\begin{align}
  & E_{1}=m^{2}\left(\bar{p}+1+ \frac{2\tilde{r}-1+2\alpha}{2n} + \frac{1-2\tilde{p}}{2m}\right)^{2}
      -\frac{1}{4}, \label{eq:1pl-e1} \\ 
  &E_{2}=m^{2}\left(\bar{p}+1 + \frac{1-2\tilde{r}+2\alpha}{2n} + \frac{2\tilde{p}-1}{2m}\right)^{2}
      -\frac{1}{4}, \label{eq:1pl-e2}  
\end{align}
and finite-dimensional unirreps associated with the structure functions
\begin{equation}
\begin{split}
  & \Phi_{1}(x) = n^{2n}m^{2m} \prod_{r=1}^{n}\left[\left(x+\frac{\tilde{r}-r}{n}\right)\left(x+
        \frac{2\alpha+\tilde{r}-r}{n}\right)\right] \\
  &\quad \times \prod_{p=1}^{m}\left[\left(\bar{p}+1-x-\frac{\tilde{p}-p}{m}\right) \left(\bar{p}+1+x+
        \frac{1-\tilde{p}-p}{m}+ \frac{2\alpha+2\tilde{r}-1}{n}\right)\right],  
\end{split}  \label{eq:1pl-uni1}
\end{equation}
\begin{equation}
\begin{split}
  &\Phi_{2}(x) =n^{2n}m^{2m}\prod_{p=1}^{m}\left[\left(x+\frac{\tilde{p}-p}{m}\right) \left(2\bar{p}+2 -x +
       \frac{2\alpha-2\tilde{r}+1}{n}+\frac{\tilde{p}+p-1}{m}\right)\right]  \\
  & \times \prod_{r=1}^{n} \left[\left(\bar{p}+1-x-\frac{\tilde{r}-r}{n}\right)\left(\bar{p}+1-x+
       \frac{2\alpha-\tilde{r}+r}{n}\right)\right].
\end{split} \label{eq:1pl-uni2}
\end{equation}
Here $\bar{p}=0$, 1, 2, \ldots, $\tilde{p} \in \{1, 2, \ldots, m\}$, and $\tilde{r} \in \{1, 2, \ldots, n\}$. 
\par
%
%
\subsection{Two-parameter Lissajous system}

The structure function given for this model by Eq.~(\ref{eq:strut-2pl}) can be factorized as
\begin{equation}
\begin{split}
  &\Phi(x,E,u) \\
  &\quad = (2n)^{4n}(2m)^{4m} \prod_{r=1}^{n}\biggl[\left(x+u-\frac{2r-1-\alpha-\beta}{2n}\right)
         \left(x+u-\frac{2r-1+\alpha+\beta}{2n}\right) \\
  &\quad\quad \times \left(x+u-\frac{2r-1+\alpha-\beta}{2n}\right)\left(x+u
         -\frac{2r-1-\alpha+\beta}{2n}\right)\biggr] \\
  &\quad\quad \times \prod_{p=1}^{2m}\left[\left(x+u-\frac{2p-1-\sqrt{1+4E}}{4m}\right)\left(x+u-
        \frac{2p-1+\sqrt{1+4E}}{4m}\right)\right].   
\end{split}
\end{equation}
\par
On using the constraints (\ref{eq:const123}), we obtain the following equivalent solutions
\begin{align}
  & u_{1}=\frac{2\tilde{r}-1+\alpha+\beta}{2n},\quad \tilde{r} \in \{1, 2, \ldots, n\},   \\
  & u_{2}=\frac{2\tilde{p}-1-\sqrt{1+4E}}{4m},\quad \tilde{p} \in \{1, 2, \ldots, 2m\},  
\end{align}
with the corresponding energy spectrum
\begin{align}
  & E_{1}=4m^{2}\left(\bar{p}+1+\frac{2\tilde{r}-1+\alpha+\beta}{2n}+\frac{1-2\tilde{p}}{4m}\right)^{2}-
       \frac{1}{4}, \label{eq:2pl-e1} \\ 
  & E_{2}=4m^{2}\left(\bar{p}+1+\frac{2\tilde{p}-1}{4m}+\frac{1-2\tilde{r}+\alpha+\beta}{2n}\right)^{2}-
       \frac{1}{4}, \label{eq:2pl-e2}
\end{align}
and the finite-dimensional unirreps associated with the structure functions
\begin{equation}
\begin{split}
  & \Phi_{1}(x) = (2n)^{4n}(2m)^{4m} \prod_{r=1}^{n}\biggl[\left(x+\frac{\tilde{r}-r+\alpha+\beta}{n}\right)  
      \left(x+\frac{\tilde{r}-r}{n}\right) \\
  & \quad \times \left(x+\frac{\tilde{r}-r+\beta}{n}\right)\left(x+\frac{\tilde{r}-r+
      \alpha}{n}\right)\biggr] \\
  & \quad \times \prod_{p=1}^{2m}\left[\left(\bar{p}+1-x-\frac{\tilde{p}-p}{2m}\right)\left(\bar{p}+1+x+
      \frac{2 \tilde{r}-1+\alpha+\beta}{n}+\frac{1-\tilde{p}-p}{2m}\right)\right], 
\end{split} \label{eq:2pl-uni1}   
\end{equation}
\begin{equation}
\begin{split}
  & \Phi_{2}(x) =(2n)^{4n}(2m)^{4m} \prod_{r=1}^{n}\biggl[\left(\bar{p}+1-x-\frac{\tilde{r}-r}{n}\right)
       \left(\bar{p}+1-x+\frac{\alpha+\beta-\tilde{r}+r}{n}\right)  \\
  &\quad \times \left(\bar{p}+1-x +\frac{\alpha-\tilde{r}+r}{n}\right)\left(\bar{p}+1-x+
       \frac{\beta-\tilde{r}+r}{n}\right)\biggr] \\
  &\quad \times \prod_{p=1}^{2m}\left[\left(x+\frac{\tilde{p}-p}{2m}\right)\left(2\bar{p}+2-x+
       \frac{\tilde{p}+p-1}{2m}+\frac{1+\alpha+\beta-2 \tilde{r}}{n}\right)\right]. 
\end{split}  \label{eq:2pl-uni2}  
\end{equation}
Here $\bar{p}=0$, 1, 2, \ldots, $\tilde{p} \in \{1, 2, \ldots, 2m\}$, and $\tilde{r} \in \{1, 2, \ldots, n\}$. 
\par
%
%
\subsection{One-step extensions of the two-parameter Lissajous system}

Similarly, the structure function given by Eq.~(\ref{eq:strut-e2pl}) can be factorized as
\begin{equation}
\begin{split}
  & \Phi(x,E,u) = (2n)^{8n}(2m)^{4m} \\
  & \quad \times \prod_{q=1}^n \biggl[\left(x+u-\frac{2q+1+\alpha-\beta-2m_1}{2n}\right)
      \left(x+u-\frac{2q-1-\alpha+\beta+2m_1}{2n}\right) \\
  & \quad \times \left(x+u-\frac{2q-1+\alpha-\beta-2m_1}{2n}\right)
      \left(x+u-\frac{2q-3-\alpha+\beta+2m_1}{2n}\right)\biggr] \\  
  & \quad \times \prod_{r=1}^n \biggl[\left(x+u-\frac{2r-1-\alpha-\beta}{2n}\right)
      \left(x+u-\frac{2r-1+\alpha+\beta}{2n}\right) \\
  & \quad \times \left(x+u-\frac{2r+1+\alpha-\beta}{2n}\right)
      \left(x+u-\frac{2r-3-\alpha+\beta}{2n}\right)\biggr] \\
  & \quad \times \prod_{p=1}^{2m} \left[\left(x+u-\frac{2p-1-\sqrt{1+4E}}{4m}\right) 
       \left(x+u-\frac{2p-1+\sqrt{1+4E}}{4m}\right)\right].   
\end{split}
\end{equation}
\par
%
%
On using the constraints (\ref{eq:const123}) to have finite-dimensional unirreps, we get the solutions
\begin{align}
  & u_{1}=\frac{2\tilde{r}-1+\alpha+\beta}{2n},\quad \tilde{r} \in \{1, 2, \ldots, n\},   \\
  & u_{2}=\frac{2\tilde{p}-1-\sqrt{1+4E}}{4m},\quad \tilde{p} \in \{1, 2, \ldots, 2m\},  
\end{align}
with the energy spectrum
\begin{align}
  & E_{1}=4m^{2}\left(\bar{p}+1+\frac{2\tilde{r}-1+\alpha+\beta}{2n}+\frac{1-2\tilde{p}}{4m}\right)^{2}-
       \frac{1}{4},  \\ 
  & E_{2}=4m^{2}\left(\bar{p}+1+\frac{2\tilde{p}-1}{4m}+\frac{1-2\tilde{r}+\alpha+\beta}{2n}\right)^{2}-
       \frac{1}{4}, 
\end{align}
and the final structure functions
\begin{equation}
\begin{split}
  & \Phi_{1}(x) = (2n)^{8n}(2m)^{4m} \\ 
  & \quad \times \prod_{q=1}^n \biggl[\left(x+\frac{\tilde{r}-q+\beta+m_1-1}{n}
      \right) \left(x+\frac{\tilde{r}-q+\alpha-m_1}{n}\right) \\
  & \quad \times \left(x+\frac{\tilde{r}-q+\beta+m_1}{n}
      \right) \left(x+\frac{\tilde{r}-q+\alpha-m_1+1}{n}\right)\biggr] \\    
  & \quad \times \prod_{r=1}^{n}\biggl[\left(x+\frac{\tilde{r}-r+\alpha+\beta}{n}\right)  
      \left(x+\frac{\tilde{r}-r}{n}\right) \\
  & \quad \times \left(x+\frac{\tilde{r}-r+\beta-1}{n}\right)\left(x+\frac{\tilde{r}-r+
      \alpha+1}{n}\right)\biggr] \\
  & \quad \times \prod_{p=1}^{2m}\left[\left(\bar{p}+1-x-\frac{\tilde{p}-p}{2m}\right)\left(\bar{p}+1+x+
      \frac{2 \tilde{r}-1+\alpha+\beta}{n}+\frac{1-\tilde{p}-p}{2m}\right)\right], 
\end{split}   
\end{equation}
\begin{equation}
\begin{split}
  & \Phi_{2}(x) =(2n)^{8n}(2m)^{4m} \\
  & \quad \times \prod_{q=1}^n \biggl[\left(\bar{p}+1-x-\frac{\tilde{r}-q-\alpha+m_1-1}{n}
      \right) \left(\bar{p}+1-x-\frac{\tilde{r}-q-\beta-m_1}{n}\right) \\
  & \quad \times \left(\bar{p}+1-x-\frac{\tilde{r}-q-\alpha+m_1}{n}
      \right) \left(\bar{p}+1-x-\frac{\tilde{r}-q-\beta-m_1+1}{n}\right)\biggr] \\ 
  & \quad \times \prod_{r=1}^{n}\biggl[\left(\bar{p}+1-x-\frac{\tilde{r}-r}{n}\right)
       \left(\bar{p}+1-x+\frac{\alpha+\beta-\tilde{r}+r}{n}\right) \\
  &\quad \times \left(\bar{p}+1-x +\frac{\alpha-\tilde{r}+r+1}{n}\right)\left(\bar{p}+1-x+
       \frac{\beta-\tilde{r}+r-1}{n}\right)\biggr] \\
  &\quad \times \prod_{p=1}^{2m}\left[\left(x+\frac{\tilde{p}-p}{2m}\right)\left(2\bar{p}+2-x+
       \frac{\tilde{p}+p-1}{2m}+\frac{1+\alpha+\beta-2 \tilde{r}}{n}\right)\right]. 
\end{split}   
\end{equation}
Here $\bar{p}=0$, 1, 2, \ldots, $\tilde{p} \in \{1, 2, \ldots, 2m\}$, and $\tilde{r} \in \{1, 2, \ldots, n\}$.
\par
%
%
\subsection{Physical spectrum}

We can compare these results with the physical spectrum of the three models provided by separation of variables. The latter is given by $E^K_{\mu}$ in Eq.~(\ref{eq:E-H-phi1}) with $K$ defined in (\ref{eq:H_theta}) and the appropriate $\epsilon_{\nu}$.\par
%
%
In the case of the one-parameter Lissajous system on the sphere, we make the following transformation
\begin{equation}
  \nu=n\nu' +a_{1},\quad a_{1} \in \{0, 1, \ldots, n-1\},
\end{equation}
\begin{equation}
  \mu=m\mu' +a_{2},\quad a_{2} \in \{0, 1, \ldots, m-1\},
\end{equation}
\begin{equation}
  \bar{p}=\nu' +\mu',
\end{equation}
use $\epsilon_{\nu}$ as given by (\ref{eq:H_phi-1}), and write the physical spectrum in the form $m^{2}( \bar{p} + \xi)-\frac{1}{4}$, where $\xi$ depends on $n,m,\alpha,a_1$, and $a_2$. By defining $a_{1}=\tilde{r}-1$ and $a_{2}=m-\tilde{p}$, the physical spectrum takes the form as given by $E_{1}$ in (\ref{eq:1pl-e1}). Alternatively, by taking $a_{2}=\tilde{p}-1$ and $a_{1}=n-\tilde{r}$, the spectrum coincides with $E_{2}$ in (\ref{eq:1pl-e2}). Thus the two solutions are different, but equivalent ways to enumerate the degenerate levels.\par
%
%
In the case of the two-paramater Lissajous system and its rational extensions, we replace simply $m$ by $2m$ everywhere and take for the physical spectrum the parameter $\epsilon_{\nu}$ as given by (\ref{eq:H_phi-2}). The spectrum obtained algebraically is thus also corroborated by the physical spectrum obtained via separation of variables for these two models.\par
%
%
\section{CONCLUSION}

In this paper, we presented an algebraic derivation of the spectrum of the one- and two-parameter Lissajous systems, which was so far unexplored. In addition, we introduced rationally extended two-parameter Lissajous systems related to Jacobi EOP of type I (or II), which are new families of superintegrable Hamiltonians with higher order integrals of motion (in fact arbitrary order) and demonstrated how the SUSYQM approach and the supercharges can be combined with ladder and shift operators to generate new superintegrable models with separation of variables in spherical coordinates. The spectrum of these models was also derived algebraically. \par
%
%
As one of the main results, we extended the Daskaloyannis approach to obtain the realizations as deformed oscillator algebras for systems separable in
spherical coordinates and for which the polynomial algebras display a more complex structure. This is achieved by exploiting previous results on a class of polynomial algebras with three generators \cite{isaac} and the presence of an extra algebraic relation that allows to write the Casimir operator in terms of the number operator $N$ and the Hamiltonian $H$, without having to compute explicitly this operator and its expression in terms of H only, and even without having to calculate all the structure constants of the polynomial algebra itself. This is a novelty that could provide a way to study algebraically many classes of superintegrable systems and have a wider applicability as many models introduced in recent years remain to be studied  algebraically using their symmetry algebra.\par
%
%
The results also showed how the studies of polynomial algebras, their Casimir operator and realizations as abstract algebraic structures in the line of Ref.~\cite{isaac} are important as a ladder and shift operators approach can be used to create the integrals and identify the polynomial algebra. The algebraic derivation of the spectrum requires an understanding of the constraints for the existence of the realizations and of the form of the Casimir operator. \par
%
%
{}Furthermore, we connected these results with the generalized Heisenberg algebra (GHA) generated by their nonpolynomial integrals of motion. This illustrates how the structure of these GHA is directly related to the additional algebraic relation and how there is a direct transformation that allows to obtain the realizations as deformed oscillator algebra and the corresponding structure function. These results highlight that the use of intermediate nonpolynomial integrals of motion that allow to obtain the polynomial ones, as discovered by Calzada, Kuru, and Negro \cite{calzada14a,calzada14b} in recent papers, is not only useful in this regard, but also for the algebraic derivation of the spectrum. \par
%
%
There are many possible generalizations of these results, in particular to the $k$-step rational extensions of these models and also to the study of systems related with Jacobi EOP of type III. The latter will require to modify the type of ladder operators used, as new levels are created below the ground state of the initial Hamiltonian in the SUSYQM approach. \par
%
%
\section*{ACKNOWLEDGMENTS}

The research of I.\ M.\ was supported by the Australian Research Council through Discovery Early Career Researcher Award DE130101067 and by a FNRS fellowship for a stay at the Universit\'e Libre de Bruxelles. He also thanks the PNTPM for its hospitality. \par
%
%
\section*{\boldmath APPENDIX A: ACTION OF $B^{\pm}_{\nu}$ AND $X^{\pm}_{\mu,\nu}$ ON WAVEFUNCTIONS FOR THE TWO-PARAMETER HAMILTONIAN AND ITS RATIONAL EXTENSION}

\renewcommand{\theequation}{A.\arabic{equation}}
\setcounter{equation}{0}

In this Appendix, we present some explicit formulas generalizing Eqs.~(\ref{eq:ladder-action-1}), (\ref{eq:X-action-1}), and (\ref{eq:X-product-1}).\par
%
%
{}For the two-parameter Hamiltonian (\ref{eq:lissajous}), we obtain the following results:
\begin{equation}
\begin{split}
  & B^+_{\nu} \Phi_{\nu} \\
  &= 4 \left(\frac{(\alpha+\beta+1+2\nu)(\nu+1)(\alpha+\beta+1+\nu)(\alpha+1+\nu)
       (\beta+1+\nu)}{\alpha+\beta+3+2\nu}\right)^{1/2}\Phi_{\nu+1}, \\
  & B^-_{\nu} \Phi_{\nu}\\
  &= 4 \left(\frac{(\alpha+\beta+1+2\nu)\nu(\alpha+\beta+\nu)(\alpha+\nu)
       (\beta+\nu)}{\alpha+\beta-1+2\nu}\right)^{1/2} \Phi_{\nu-1},
\end{split} 
\end{equation}
\begin{equation}
\begin{split}
  & X^+_{\mu,\nu} \Psi^{(K)}_{\mu,\nu} \\
  &= 2^{2n} \left(\frac{(\alpha+\beta+1+2\nu)(\nu+n)!\Gamma(\alpha+\beta+\nu+1+n)
      \Gamma(\alpha+\nu+1+n)}{(\alpha+\beta+1+2\nu+2n)\nu!
      \Gamma(\alpha+\beta+\nu+1)\Gamma(\alpha+\nu+1)}\right)^{1/2} \\
  & \quad \times \left(\frac{\Gamma(\beta+\nu+1+n)\mu!\Gamma(\mu+2K+1+2m)}{\Gamma(\beta+\nu+1)
      (\mu-2m)!\Gamma(\mu+2K+1)}\right)^{1/2} \Psi^{(K+2m)}_{\mu-2m,\nu+n}, \\
  & X^-_{\mu,\nu} \Psi^{(K)}_{\mu,\nu} \\
  &= 2^{2n} \left(\frac{(\alpha+\beta+1+2\nu)\nu!\Gamma(\alpha+\beta+\nu+1)
      \Gamma(\alpha+\nu+1)}{(\alpha+\beta+1+2\nu-2n)(\nu-n)!
      \Gamma(\alpha+\beta+\nu+1-n)\Gamma(\alpha+\nu+1-n)}\right)^{1/2} \\
  & \quad \times \left(\frac{\Gamma(\beta+\nu+1)(\mu+2m)!\Gamma(\mu+2K+1)}{\Gamma(\beta+\nu+1-n)
      \mu!\Gamma(\mu+2K+1-2m)}\right)^{1/2} \Psi^{(K-2m)}_{\mu+2m,\nu-n},
\end{split}
\end{equation}
\begin{equation}
\begin{split}
  & X^+_{\mu+2m,\nu-n} X^-_{\mu,\nu} \Psi^{(K)}_{\mu,\nu} \\
  &= 2^{4n} \frac{\nu!\Gamma(\alpha+\beta+\nu+1)\Gamma(\alpha+\nu+1)\Gamma(\beta+\nu+1)(\mu+2m)!}
        {(\nu-n)!\Gamma(\alpha+\beta+\nu+1-n)\Gamma(\alpha+\nu+1-n)\Gamma(\beta+\nu+1-n)\mu!} \\
  &\quad \times \frac{\Gamma(\mu+2K+1)}{\Gamma(\mu+2K+1-2m)} \Psi^{(K)}_{\mu,\nu}, \\
  & X^-_{\mu-2m,\nu+n} X^+_{\mu,\nu} \Psi^{(K)}_{\mu,\nu} \\
  &= 2^{4n} \frac{(\nu+n)!\Gamma(\alpha+\beta+\nu+1+n)\Gamma(\alpha+\nu+1+n)\Gamma(\beta+\nu+1+n)
        \mu!}{\nu!\Gamma(\alpha+\beta+\nu+1)\Gamma(\alpha+\nu+1)\Gamma(\beta+\nu+1)(\mu-2m)!} \\
  &\quad \times \frac{\Gamma(\mu+2K+1+2m)}{\Gamma(\mu+2K+1)} \Psi^{(K)}_{\mu,\nu}.
\end{split} \label{eq:X-product-2}
\end{equation}
\par
%
%
Similarly, calculations for its rational extension (\ref{eq:ext-lissajous}) yield
\begin{equation}
\begin{split}
  & B^+_{\nu} \Phi_{\nu} \\
  &= 16 [(\alpha+\nu-m_1+1)(\alpha+\nu-m_1+2)(\beta+\nu+m_1)(\beta+\nu+m_1+1)]^{1/2} \\
  &\quad \times \left(\frac{(\alpha+\beta+1+2\nu)(\nu+1)(\alpha+\beta+1+\nu)(\alpha+2+\nu)
       (\beta+\nu)}{\alpha+\beta+3+2\nu}\right)^{1/2}\Phi_{\nu+1}, \\
  & B^-_{\nu} \Phi_{\nu}\\
  &= 16 [(\alpha+\nu-m_1+1)(\alpha+\nu-m_1)(\beta+\nu+m_1)(\beta+\nu+m_1-1)]^{1/2} \\
  &\quad \times \left(\frac{(\alpha+\beta+1+2\nu)\nu(\alpha+\beta+\nu)(\alpha+\nu+1)
       (\beta+\nu-1)}{\alpha+\beta-1+2\nu}\right)^{1/2} \Phi_{\nu-1},
\end{split} 
\end{equation}
\begin{equation}
\begin{split}
  & X^+_{\mu,\nu} \Psi^{(K)}_{\mu,\nu} \\
  &= 2^{4n} \frac{\Gamma(\alpha+\nu-m_1+n+1)\Gamma(\beta+\nu+m_1+n)}{\Gamma(\alpha+\nu-m_1+2)
      \Gamma(\beta+\nu+m_1+1)} \\
  &\quad \times [(\alpha+\nu-m_1+1)(\alpha+\nu-m_1+n+1)(\beta+\nu+m_1)(\beta+\nu+m_1+n)]^{1/2} \\
  &\quad \times \left(\frac{(\alpha+\beta+1+2\nu)(\nu+n)!\Gamma(\alpha+\beta+\nu+1+n)
      \Gamma(\alpha+\nu+2+n)}{(\alpha+\beta+1+2\nu+2n)\nu!
      \Gamma(\alpha+\beta+\nu+1)\Gamma(\alpha+\nu+2)}\right)^{1/2} \\
  & \quad \times \left(\frac{\Gamma(\beta+\nu+n)\mu!\Gamma(\mu+2K+1+2m)}{\Gamma(\beta+\nu)
      (\mu-2m)!\Gamma(\mu+2K+1)}\right)^{1/2} \Psi^{(K+2m)}_{\mu-2m,\nu+n}, \\
  & X^-_{\mu,\nu} \Psi^{(K)}_{\mu,\nu} \\
  &= 2^{4n} \frac{\Gamma(\alpha+\nu-m_1+1)\Gamma(\beta+\nu+m_1)}{\Gamma(\alpha+\nu-m_1-n+2)
      \Gamma(\beta+\nu+m_1-n+1)} \\
  &\quad \times [(\alpha+\nu-m_1+1)(\alpha+\nu-m_1-n+1)(\beta+\nu+m_1)(\beta+\nu+m_1-n)]^{1/2} \\
  &\quad \times \left(\frac{(\alpha+\beta+1+2\nu)\nu!\Gamma(\alpha+\beta+\nu+1)
      \Gamma(\alpha+\nu+2)}{(\alpha+\beta+1+2\nu-2n)(\nu-n)!
      \Gamma(\alpha+\beta+\nu+1-n)\Gamma(\alpha+\nu+2-n)}\right)^{1/2} \\
  & \quad \times \left(\frac{\Gamma(\beta+\nu)(\mu+2m)!\Gamma(\mu+2K+1)}{\Gamma(\beta+\nu-n)
      \mu!\Gamma(\mu+2K+1-2m)}\right)^{1/2} \Psi^{(K-2m)}_{\mu+2m,\nu-n},
\end{split}
\end{equation}
\begin{equation}
\begin{split}
  & X^+_{\mu+2m,\nu-n} X^-_{\mu,\nu} \Psi^{(K)}_{\mu,\nu} \\
  &= 2^{8n} \frac{\Gamma(\alpha+\nu-m_1+1)\Gamma(\alpha+\nu-m_1+2)\Gamma(\beta+\nu+m_1)
        }{\Gamma(\alpha+\nu-m_1-n+1)\Gamma(\alpha+\nu-m_1-n+2)\Gamma(\beta+\nu+m_1-n)} \\ 
  &\quad \times \frac{\Gamma(\beta+\nu+m_1+1)\nu!\Gamma(\alpha+\beta+\nu+1)
        \Gamma(\alpha+\nu+2)}{\Gamma(\beta+\nu+m_1-n+1)(\nu-n)!
        \Gamma(\alpha+\beta+\nu+1-n)\Gamma(\alpha+\nu+2-n)} \\
  &\quad \times \frac{\Gamma(\beta+\nu)(\mu+2m)!\Gamma(\mu+2K+1)}{\Gamma(\beta+\nu-n)\mu!
        \Gamma(\mu+2K+1-2m)} \Psi^{(K)}_{\mu,\nu}, \\
  & X^-_{\mu-2m,\nu+n} X^+_{\mu,\nu} \Psi^{(K)}_{\mu,\nu} \\
  &= 2^{8n} \frac{\Gamma(\alpha+\nu-m_1+n+1)\Gamma(\alpha+\nu-m_1+n+2)\Gamma(\beta+\nu+m_1+n)
        }{\Gamma(\alpha+\nu-m_1+1)\Gamma(\alpha+\nu-m_1+2)\Gamma(\beta+\nu+m_1)} \\ 
  &\quad \times \frac{\Gamma(\beta+\nu+m_1+n+1)(\nu+n)!\Gamma(\alpha+\beta+\nu+n+1)
        \Gamma(\alpha+\nu+n+2)}{\Gamma(\beta+\nu+m_1+1)\nu!
        \Gamma(\alpha+\beta+\nu+1)\Gamma(\alpha+\nu+2)} \\
  &\quad \times \frac{\Gamma(\beta+\nu+n)\mu!\Gamma(\mu+2K+1+2m)}{\Gamma(\beta+\nu)(\mu-2m)!
        \Gamma(\mu+2K+1)} \Psi^{(K)}_{\mu,\nu}.
\end{split} \label{eq:X-product-3}
\end{equation}
\par
\par
%
%

\end{document}